\documentclass[aps,prl,superscriptaddress, twocolumn, amsmath,amssymb,floatfix]{revtex4-1} 
\usepackage{graphicx}
\usepackage{bm,siunitx, balance, color,soul}
\usepackage[colorlinks=True, allcolors=blue]{hyperref}

\begin{document}
    \title{Glass formation in binary alloys with different atomic symmetries}

    \author{Yuan-Chao Hu}
    \affiliation{Department of Mechanical Engineering \& Materials Science, Yale University, New Haven, Connecticut 06520, USA}

   \author{Kai Zhang}
   \affiliation{Division of Natural and Applied Sciences, Duke Kunshan University, Kunshan, Jiangsu, 215300, China}

    \author{Sebastian A. Kube}
    \affiliation{Department of Mechanical Engineering \& Materials Science, Yale University, New Haven, Connecticut 06520, USA}
    
    \author{Jan Schroers}
    \affiliation{Department of Mechanical Engineering \& Materials Science, Yale University, New Haven, Connecticut 06520, USA}

    \author{Mark D. Shattuck}
    \affiliation{Benjamin Levich Institute and Physics Department, The City College of New York, New York, New York 10031, USA.}

    \author{Corey S. O'Hern}
    \email[]{corey.ohern@yale.edu}
    \affiliation{Department of Mechanical Engineering \& Materials Science, Yale University, New Haven, Connecticut 06520, USA}
    \affiliation{Department of Physics, Yale University, New Haven, Connecticut 06520, USA.}
    \affiliation{Department of Applied Physics, Yale University, New Haven, Connecticut 06520, USA.}
    \date{\today}

\begin{abstract}

Prediction of the glass forming ability (GFA) of alloys remains a
major challenge. We are not able to predict the composition dependence
of the GFA of even binary alloys. To investigate the effect of each
element's propensity to form particular crystal structures on glass
formation, we focus on binary alloys composed of elements with the
same size, but different atomic symmetries using the patchy-particle
model. For mixtures with atomic symmetries that promote different
crystal structures, the minimum critical cooling rate $R_c$ is only a
factor of $5$ lower than that for the pure substances. For mixtures
with different atomic symmetries that promote local crystalline and
icosahedral order, the minimum $R_c$ is more than $3$ orders of
magnitude lower than that for pure substances. Results for $R_c$ for
the patchy-particle model are in agreement with those from embedded
atom method simulations and sputtering experiments of NiCu, TiAl, and high
entropy alloys.
\end{abstract}

\maketitle

Bulk metallic glasses (BMGs), which are multi-component alloys with
disordered atomic-scale structure, are a promising materials class
because they combine metal-like strength with plastic-like
processability~\cite{kumar_nanomoulding_2009,chen_mechanical_2008}. Despite
their potential, they have not been widely used,
likely because current BMGs do not combine multiple advantageous
properties, such as high strength, high fracture toughness, and low
material
cost~\cite{johnson_is_2015,wang_bulk_2004,sun_thermomechanical_2016,li_how_2017}.

A first step in the BMG design process is the ability to predict the
glass-forming ability (GFA), or critical cooling rate $R_c$ below
which crystallization occurs. BMGs with good GFA, e.g. $\rm
Pd_{42.5}Cu_{30}Ni_{7.5}P_{20}$ with $R_c \sim 10^{-2}$~K/s, have been
identified mainly through time-consuming experiments that are guided
by empirical
rules~\cite{nishiyama_glass-forming_2002,inoue_stabilization_2000}. The
number of alloys that can potentially form metallic glasses is
enormous, {\it i.e.} more than $10^{6}$ for four-component
alloys with $32$ possible elements and $1\%$ increments in composition
of the four elements~\cite{li_how_2017}.  However, even using
the latest high-throughput sputtering techniques, researchers can only
characterize a minute fraction of
these~\cite{ding_combinatorial_2014,bordeenithikasem_determination_2017,ren_accelerated_2018,li_high-temperature_2019}.

We seek to develop a computational platform to predict the GFA of
alloys. We focus on binary systems (with elements A and B) and
determine whether the best GFA occurs for equal proportions of A and
B, or for the A- or B-rich systems. The answer to even this simple
question is unknown for most binary alloys. Prior studies have focused
on the role of atomic size and cohesive energy in determining the GFA
of alloys~\cite{laws_predictive_2015,kai_zhang_computational_2013,hu_tuning_2019}. Other studies suggest that the composition with the best GFA
can be predicted from the equilibrium liquidus
curve~\cite{turnbull_under_1969,johnson_quantifying_2016}. However,
many alloys do not possess eutectic points, and there are numerous
examples where the composition with the best GFA deviates from the deepest
eutectic~\cite{xia_binary_2006,tan_optimum_2003,wang_bulk_2004}. Few
studies have considered the effect of each element's propensity to
form particular crystal structures on the GFA of alloys.

\begin{figure}[t!]
\centering
\includegraphics[width = \linewidth]{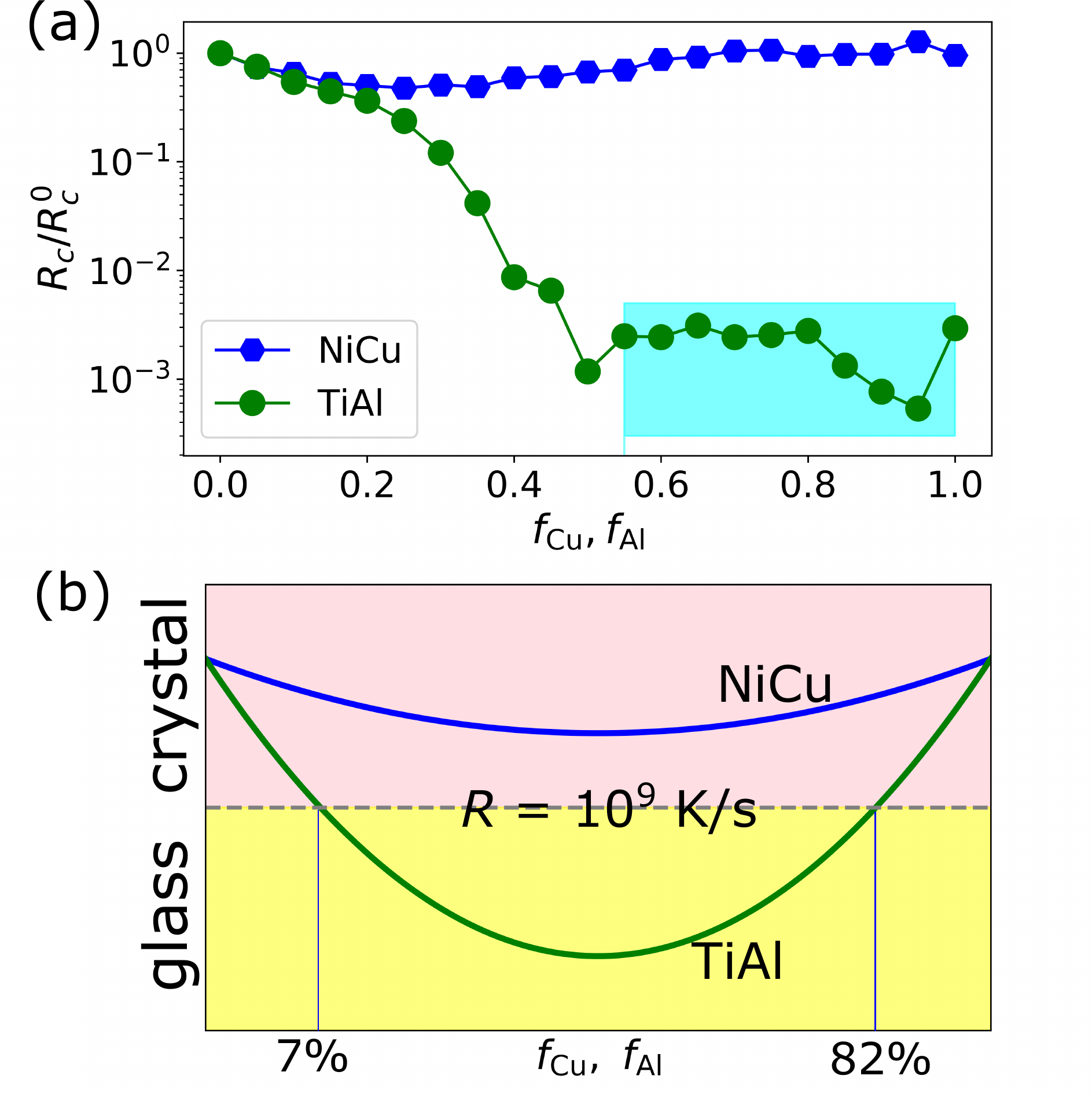}
\caption{(a) The critical cooling rate $R_c/R_c^0$ for NiCu (normalized by $R_c^0$ for pure Ni) versus the fraction $f_{\rm Cu}$ of Cu atoms, and TiAl (normalized by $R_c^0$ for pure Ti) versus the fraction $f_{\rm Al}$ of Al atoms, 
obtained using EAM simulations. The binary alloys with $f_{\rm Al} \gtrsim 0.5$ (cyan region) form quasicrystals for $R<R_c$. 
(b) Schematic diagram of solidification for NiCu and TiAl based on co-sputtering experiments, which correspond to $R \sim 10^9$~K/s. NiCu alloys crystallize over the full range of $f_{\rm Cu}$, while TiAl alloys form glasses for $0.07 < f_{\rm Al} < 0.82$.}
\label{fig1}
\end{figure}

To simplify the problem, we consider binary alloys for which the
atomic radii are the same, and investigate how the GFA depends on the
competing crystalline phases of the pure substances. We first
investigated the GFA of two specific binary alloys, NiCu and TiAl,
using MD simulations of embedded atom method (EAM) potentials to
determine $R_c$ versus alloy
composition~\cite{onat_optimized_2013,zope_interatomic_2003}. When the
pure substances crystallize, Ni and Cu form face-centered cubic (FCC)
crystals; Ti forms hexagonal close packed (HCP) and Al forms FCC
crystals in equilibrium.  (Details of the EAM simulations are provided
in the Supplemental Materials (SM)~\cite{PRSI}.) We show in
Fig.~\ref{fig1} (a) that $R_c$ for NiCu varies by less than a factor
of $5$ over the full range of composition. In contrast, $R_c$ for TiAl
decreases by more than three orders of magnitude as the fraction of Al
is increased.  We find similar results for the GFA of NiCu and TiAl
alloys in co-sputtering experiments, which correspond
to $R \sim 10^9$ K/s (Fig.~\ref{fig1} (b)). (See SM for experimental
details.) We observe only crystallized samples for NiCu over the full
range of compositions, whereas there is a wide range of compositions
where amorphous samples occur for TiAl. Although we have not
determined the compositions with the best GFA in these two alloys,
Fig.~\ref{fig1} (b) demonstrates the large difference in their
GFA. 

Although Al crystallizes into FCC structures in equilibrium,
experimental studies have shown that Al-based BMGs possess local
icosahedral order centered on the Al atoms and form metastable
quasicrystals~\cite{inoue_amorphous_1998,shechtman_metallic_1984,chen_calorimetric_1988}. In
addition, EAM simulations have shown that pure Al forms quasicrystals
by rapid quenching~\cite{prokhoda2014quasicrystals}. The above results
for TiAl alloys suggest that mixtures of elements with crystalline and
icosahedral (ICO) atomic symmetries, and similar atomic sizes, can
yield alloys with $R_c$ that are several orders of magnitude lower
than that for pure systems.

A limitation of EAM potentials~\cite{cheng_atomic-level_2011}
is that the atomic symmetry of the elements cannot be tuned
independently, while keeping other important features, such as
atomic size and cohesive energy, fixed.  To overcome this limitation,
we perform MD simulations of the patchy-particle
model~\cite{zhang_glass-forming_2015} for binary alloys, where small
patches on the surfaces of the same types of atoms attract
each other when they are aligned (and atoms of different types
interact via the Lennard-Jones potential). (See SM.) Using
this model, we study the GFA of binary mixtures of the same-sized
atoms with different atomic symmetries ({\it e.g.} BCC, FCC, and HCP).
Systems that contain atoms with a given symmetry crystallize with
that particular symmetry at low cooling rates. In addition, we study
mixtures of atoms with crystalline and ICO symmetries by controlling
the number and placement of patches on the atom surfaces.

We find that the GFA of binary alloys modeled using the
patchy-particle interaction possesses a minimum within $0 < f_B
< 1$, whose location depends on the atomic symmetry and cohesive
energies of the elements. In contrast, the melting temperature
$T_m$ of the crystalline solids varies approximately linearly with
composition. We find that the composition with the best GFA
corresponds to that for which the local icosohedral order in the
liquid state is maximized. However, if the icosohedral ordering is too
strong, metastable quasicrystals form, which decreases the
GFA. Moreover, $R_c$ for mixtures of atoms with different atomic
symmetries and cohesive energies can be collapsed by the amount of
local icosohedral order in the system.

\begin{figure}[b!]
\includegraphics[width = 0.5\textwidth]{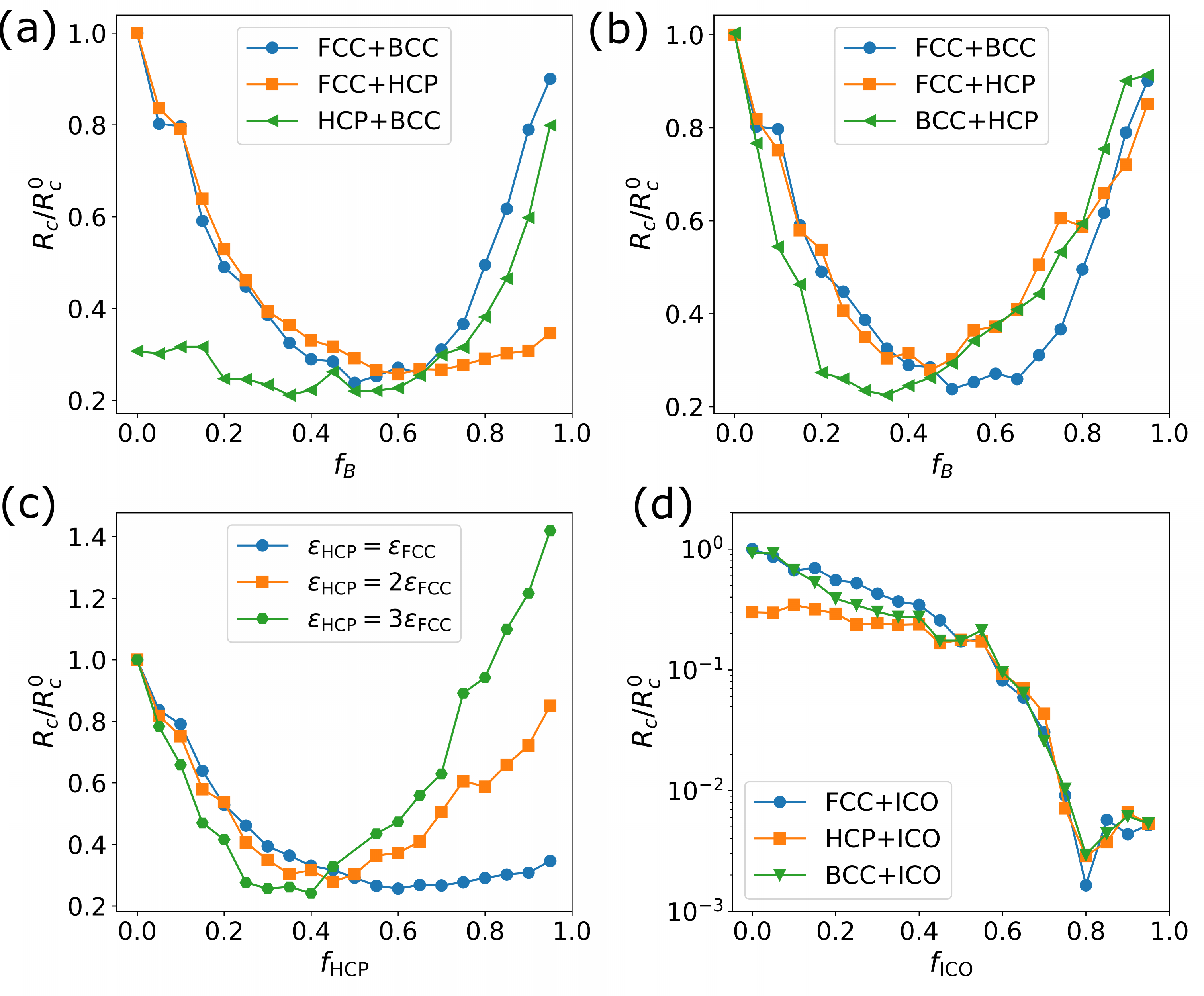}
\caption{$R_c$ for binary alloys (normalized by $R_c^0$ for the pure system with 
FCC symmetry) using the patchy-particle model. (a) $R_c/R_c^0$ for binary mixtures with $A$- and $B$-atoms [(A: FCC, B: BCC; circles), (A: FCC, B: HCP; squares), and (A: HCP, B: BCC; triangles)] versus the fraction of $B$ atoms $f_B$ with $\epsilon_{\rm BCC}/\epsilon_{\rm FCC}=\epsilon_{\rm HCP}/\epsilon_{\rm FCC}=1.0$. (b) $R_c/R_c^0$ for binary mixtures, where the pure substances (with FCC, BCC, or HCP symmetries) have similar $R_c$. We set $\epsilon_{\rm BCC}/\epsilon_{\rm FCC}=1.0$ and $\epsilon_{\rm HCP}/\epsilon_{\rm FCC}=2.0$. (c) $R_c/R_c^0$ for binary mixtures with FCC and HCP symmetries, and $\epsilon_{\rm HCP}/\epsilon_{\rm FCC}=1.0$, $2.0$, and $3.0$. (d) $R_c/R_c^0$ for binary mixtures of atoms with crystalline and icosahedral symmetries and the same cohesive energies.}
\label{fig2}
\end{figure}

In Fig.~\ref{fig2}, we show $R_c$ for binary alloys using the patchy particle model. To measure $R_c$, we 
cool the alloys linearly from the liquid state to zero temperature 
at rate $R$ and define $R_c$ as the rate below which the zero-temperature system develops strong bond orientational order. (See SM.) In (a), we consider 
three binary
alloys with FCC-BCC, FCC-HCP, and HCP-BCC symmetries for elements $A$-$B$ and the same cohesive 
energies $\epsilon_{AA}=\epsilon_{BB}$. Pure substances with HCP 
symmetry have the lowest $R_c$, while $R_c$ is similar for pure substances with 
FCC and BCC symmetries.  In general, we find that $R_c$ is minimal for non-pure 
substances. For
FCC-BCC binary alloys, the composition with the best GFA has $f_B \approx 0.5$. In contrast, for
binary alloys containing atoms with HCP symmetry, the system with minimum $R_c$ has a majority 
of HCP atoms. 

In Fig.~\ref{fig2} (b), we plot $R_c$ for binary alloys containing atoms 
that have FCC, BCC, and HCP symmetries, but the pure substances have similar GFA (by varying the cohesive energies). 
As in Fig.~\ref{fig2} (a), $R_c$
possesses a minimum in the range $0 < f_B < 1$. For binary alloys containing atoms with BCC symmetry, 
the system with the lowest $R_c$ has a majority of BCC atoms. For binary alloys with atoms with FCC and HCP 
symmetries, $f_B \approx 0.5$ has the best GFA since FCC and HCP crystal structures are similar.  

In Fig.~\ref{fig2} (c), we show $R_c$ for binary alloys containing
atoms with FCC and HCP symmetries versus the HCP-fraction
$f_{\rm HCP}$, for three cases where HCP crystals have different GFAs
(by adjusting $\epsilon_{\rm HCP}/\epsilon_{\rm FCC}$). We find that
as $R_c$ at $f_{\rm HCP}=1$ decreases, $f_{\rm HCP}$ with the best GFA
increases.  These results emphasize that the location of the minimum
in $R_c$ is influenced by the GFA of the
pure substances, which depends on their atomic symmetry and cohesive
energy.

As shown in Fig.~\ref{fig2}, for binary alloys containing same-sized
atoms, but different crystalline symmetries, the minimum $R_c$ changes
by only a factor of $5$ relative to that for the pure substances. For
binary alloys with elements of the same atomic sizes {\it and}
symmetries, we showed previously that $R_c$ scales with the ratio of
the cohesive energies of the pure
substances~\cite{zhang_glass-forming_2015,hu_tuning_2019}. Thus,
results for $R_c$ for the patchy-particle model are in general
agreement with those for EAM simulations of NiCu (with
$\epsilon_{\rm Ni}/\epsilon_{\rm Cu} \approx 1.3$) in Fig.~\ref{fig1}
(a), as well as experimental studies of mixtures of Ar and Kr (with
$\epsilon_{\rm Kr}/\epsilon_{\rm Ar} \approx
1.45$)~\cite{schottelius_crystal_2020}.

Motivated by the results for EAM simulations of TiAl in
Fig.~\ref{fig1} (a), we show $R_c$ for binary alloys containing atoms
with ICO and different crystalline symmetries in Fig.~\ref{fig2}
(d). $R_c$ decreases modestly (by less than an order of magnitude) for
$f_{\rm ICO} \lesssim 0.5$, and decreases dramatically (by more than
two orders of magnitude) for $0.5 \lesssim f_{\rm ICO} \lesssim
0.8$. When $f_{\rm ICO} \gtrsim 0.8$, the system can form
quasicrystals~\cite{keys_how_2007}, which causes $R_c$ to increase as
$f_{\rm ICO} \rightarrow 1$. (See SM for methods to detect
quasicrystals.) Note that $R_c$ for elements with ICO
symmetry is much lower than that for elements with crystalline
symmetry. We find that $R_c(f_{\rm ICO})$ possesses a minimum near
$f_{\rm ICO} \sim 0.8$. The non-monotonic behavior of $R_c(f_{\rm
  ICO})$ can be rationalized by considering the interfacial free
energy barrier for crystal
nucleation~\cite{tanaka_roles_2003,tanaka_relationship_2005,shen_icosahedral_2009,keys_how_2007}.
In the crystal-forming regime with $f_{\rm ICO} \lesssim 0.8$, local
icosahedral order is incompatible with crystalline symmetry, and thus
increasing $f_{\rm ICO}$ enhances the free energy barrier for crystal
nucleation, leading to decreases in $R_c$.  However, for $f_{\rm ICO}
\gtrsim 0.8$, ICO symmetry becomes compatible with quasicrystalline
order, reducing the interfacial free energy barrier and increasing
$R_c$.

\begin{figure}[t!]
\includegraphics[width = \linewidth]{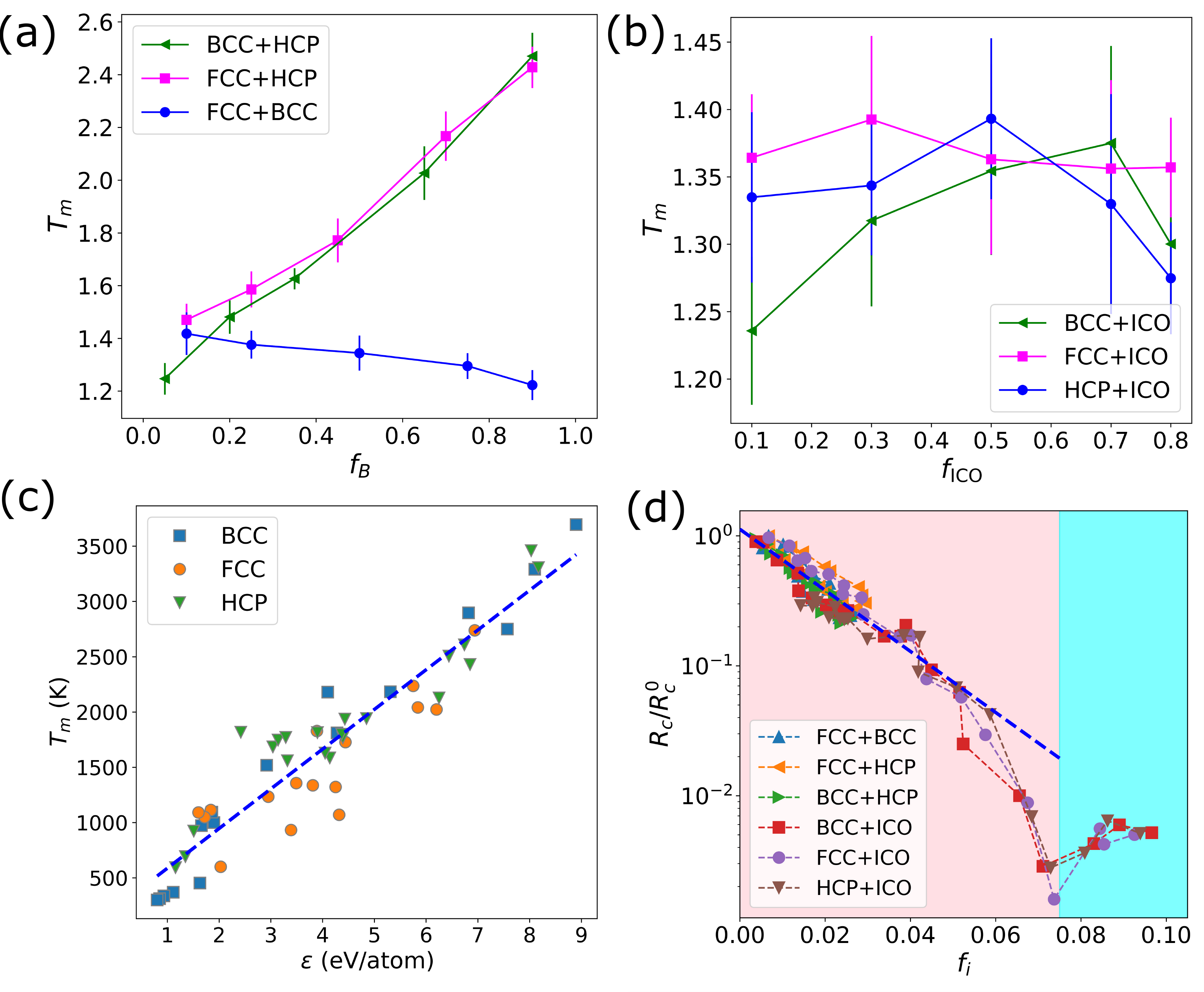}
\caption{(a) Melting temperature $T_m$ for binary alloys (in units of $\epsilon_{AA}/k_B$) using the patchy-particle 
model for mixtures of A- and B- atoms. The pure substances have similar $R_c$; see Fig.~\ref{fig2} (b). 
(b) $T_m$ for binary mixtures of atoms with crystalline and ICO symmetries. $T_m$ in (a) and (b) are obtained by heating the quenched crystalline solids to high temperature at rates $R_h \sim R_c^0$. (c) $T_m$ versus the cohesive energy per particle $\epsilon$ for $53$ pure metals with BCC, FCC, and HCP symmetries in their equilibrium solid forms. The blue dashed line gives $T_m=0.03\epsilon/k_B$, where $k_B$ is the Boltzmann constant.
(d) $R_c$ normalized by $R_c^0$ for pure substances with FCC symmetry versus the fraction of atoms $f_i$ with local icosahedral order in binary alloys using the patchy-particle model. $f_i$ is measured 
at zero temperature using the lowest $R$ at which all systems remain disordered. For $f_i \lesssim 0.075$, FCC, BCC, and HCP structures form for $R < R_c$. In the cyan region, systems 
form quasicrystals for $R < R_c$. The blue dashed line indicates exponential decay, $R_c/R_c^0 \sim \exp(-23.5f_i)$.}
\label{fig3}
\end{figure}

Prior studies suggest that the melting temperature $T_m$ of
alloys can be used to predict
$R_c$~\cite{johnson_quantifying_2016}. To test this hypothesis, we
measured $T_m$ for all binary mixtures in
Fig.~\ref{fig2}~\cite{lee_criteria_2003}. In Fig.~\ref{fig3} (a), we
show $T_m$ for binary alloys containing atoms with different
crystalline symmetries, where the pure substances have the same
GFA. From experimental data in Fig.~\ref{fig3} (c), $T_m$ for pure
substances scales roughly linearly with the cohesive energy, although
the atomic symmetry gives rise to
deviations~\cite{halpern_dimer_2012,guinea_scaling_1984}.  Thus, $T_m$
for binary alloys containing atoms with different crystalline
symmetries is roughly linear in $f_B$, and the sign of the slope is
determined by the sign of $\epsilon_{BB} -\epsilon_{AA}$. We contrast
this behavior for $T_m(f_B)$ with that for $R_c(f_B)$, which possesses
a minimum in the range $0 < f_B < 1$. In Fig.~\ref{fig3} (b), we show
$T_m$ for binary alloys containing atoms with ICO and crystalline
symmetries. In this case, $T_m$ is nearly constant for $f_{\rm ICO}
\gtrsim 0.5$, whereas $R_c$ decreases by more than $2$ orders of
magnitude. Thus, we do not find a strong correlation between $T_m$ and
GFA in our model binary alloys.

Several studies have characterized the local structural order,
such as the size and shape of Voronoi polyhedra, local bond
orientational order, and changes of nearest neighbor atoms, in
glass-forming materials as they are
cooled~\cite{cheng_atomic-level_2011}. In particular, researchers have
found that the number of atoms with local icosahedral order increases
when good glass-formers are cooled toward the glass
transition~\cite{cheng_atomic_2009}.  Thus, one suggestion for
improving the GFA is to maximize local icosahedral order. In
Fig.~\ref{fig3}(d), we show that $R_c$ for all of the patchy-particle
systems studied collapses when plotted against the fraction $f_i$ of
atoms in the system that have local icosahedral order, where the
icosohedral order is characterized using rapid quenches for which all
of the systems remain disordered. (See SM for the definition of local
icosohedral order.) $R_c(f_i)$ has several key features. First, for
$f_i \lesssim 0.06$, where most of the data for the binary mixtures
containing atoms with crystalline symmetries exists, $R_c$ decays
exponentially with increasing $f_i$. In the regime $0.06 \lesssim f_i
\lesssim 0.075$, $R_c$ decreases more rapidly. For $f_i \gtrsim
0.075$, since the system can form quasicrystals, $R_c$
begins to increase. Thus, we predict non-monotonic behavior in
$R_c(f_i)$.  

\begin{figure}[b!]
\includegraphics[width = \linewidth]{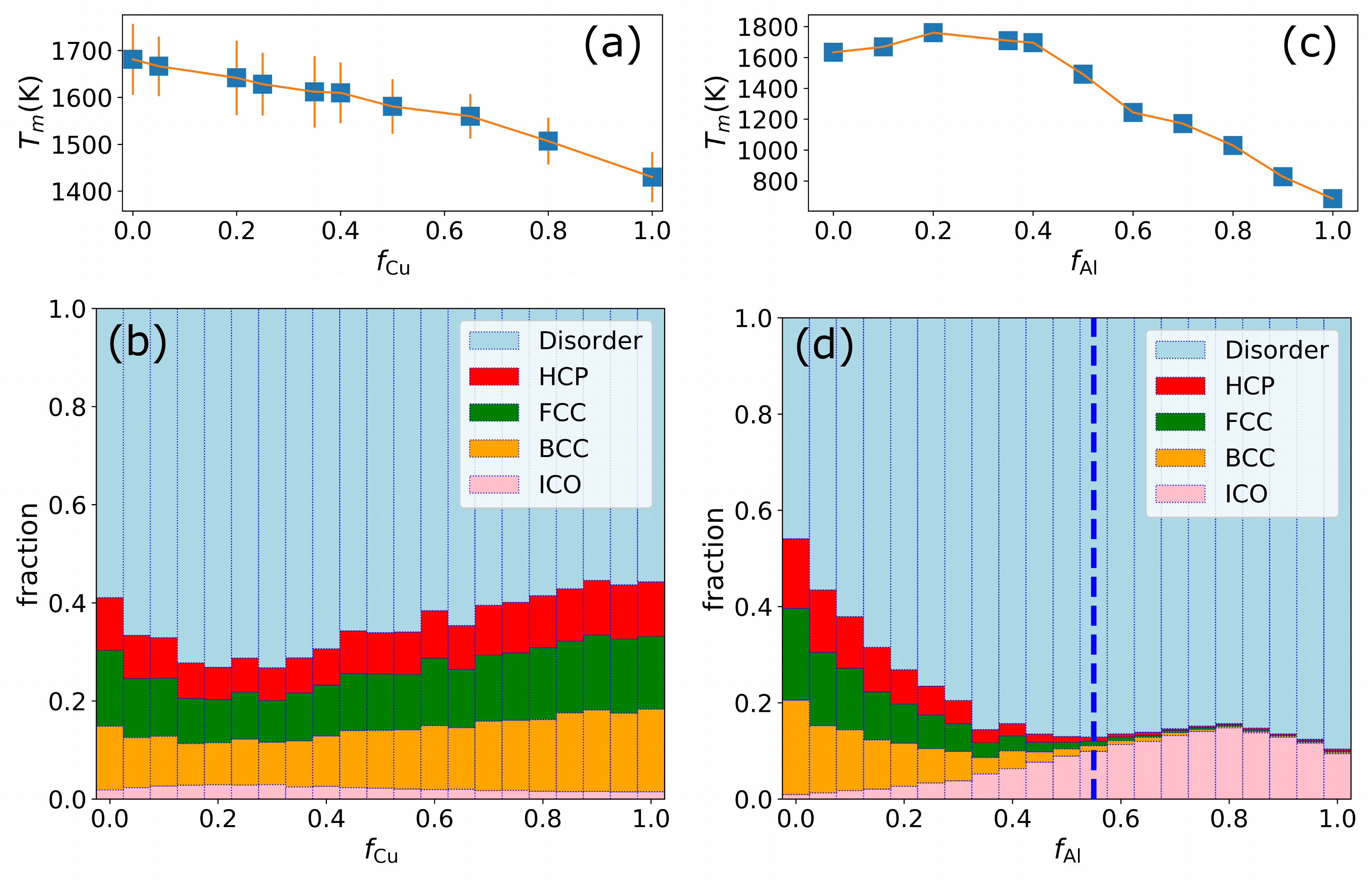}
\caption{(a) $T_m$ versus $f_{\rm Cu}$ for EAM simulations of NiCu using $R_h=10^{11}$ K/s. (b) Fraction of atoms with a given local order: HCP, FCC, BCC, ICO, or other disordered motifs versus $f_{\rm Cu}$ for zero-temperature systems at $R > R_c$. (c) $T_m$ versus $f_{\rm Al}$ for EAM simulations of TiAl using $R_h=10^{10}$ K/s. (d) Fraction of atoms with a given local order: HCP, FCC, BCC, ICO, and other disordered motifs versus $f_{\rm Al}$ at $R>R_c$.  For $f_{\rm Al} > 0.5$ (vertical dashed line), quasicrystals form for $R<R_c$. The local order in (b) and (d) is measured at $R \sim 10^{13}$~K/s.}
\label{fig4}
\end{figure}

To what extent are the results for the patchy-particle model consistent with those for the EAM simulations of NiCu and TiAl?  First, in Fig.~\ref{fig4} (a) and (c), we show $T_m$ versus $f_{\rm Cu}$ for 
NiCu and versus $f_{\rm Al}$ for TiAl alloys, which are consistent with the experimental melting curves~\cite{murray1986binary}. For NiCu, $T_m$ 
decreases roughly linearly from $\sim 1700$~K to $\sim 1400$~K over the range $0 < f_{\rm Cu} < 1$.  In contrast, $R_c(f_{\rm Cu})$ for NiCu possesses a shallow minimum near $f_{\rm Cu} \sim 0.25$. For TiAl, $T_m$ has a small maximum at $\sim 1800$~K for $f_{\rm Al} \sim 0.3$, and then $T_m$ decreases monotonically for $f_{\rm Al} \gtrsim 0.3$.  In contrast, $R_c(f_{\rm Al})$ decreases over the range $0 < f_{\rm Al} < 0.5$ and has a minimum for $f_{\rm Al} \sim 0.9$-$0.95$ (although the precise location of the minimum is affected by the degree of quasicrystalline order). These results further emphasize the decoupling of $T_m$ and $R_c$. 
Importantly, as shown in Fig.~\ref{fig4} (b) for NiCu and (d) for TiAl, the composition region with the best GFA is the same as that with the 
largest fraction of atoms with icosahedral order, and a minimal amount of (FCC, HCP, and BCC) crystalline order. 

\begin{figure}[t!]
\includegraphics[width = \linewidth]{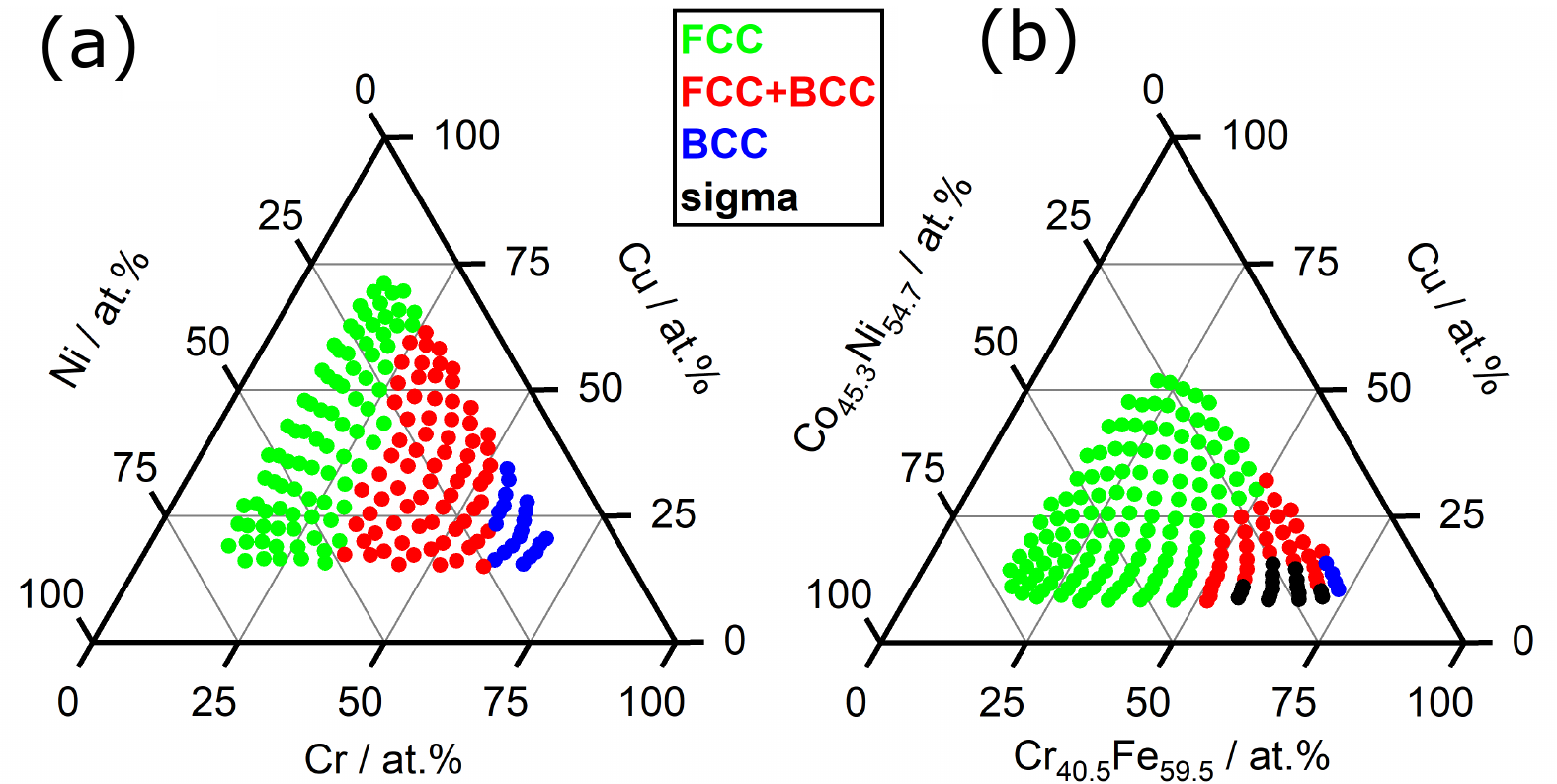}
\caption{Crystallization of sputtered (a) CrNiCu and (b) CrFe-CoNi-Cu alloys. In (a), systems along the binary NiCu axis form FCC crystals, in agreement with the pure substances. With increasing fraction of Cr (with BCC symmetry), the structure transitions to BCC. Crystal formation over the full composition range indicates that $R_c >10^9$ K/s for all CrNiCu (and NiCu) alloys in experiments. Similarly, in (b), we find crystal formation over the full range of compositions in CrFe-CoNi-Cu alloys~\cite{kube_phase_2019}, despite several competing crystalline phases.}
\label{fig5}
\end{figure}

Additional results from co-sputtering experiments~\cite{kube_phase_2019} on multi-component alloys with same-sized atoms provide further support for our findings. (See SM.) As shown in Fig.~\ref{fig5} (a), all compositions for NiCuCr (and NiCu) alloys crystallize for $R \sim 10^9$ K/s.  We also show results in Fig.~\ref{fig5} (b) for the quinary alloy CrFe-CoNi-Cu. All compositions crystallize, despite the fact that the individual elements form different crystalline phases, confirming our results for the patchy-particle model for binary alloys without ICO symmetry. 

In summary, we employed MD simulations of EAM potentials and the
patchy-particle model to investigate the influence of atomic
symmetry on the GFA of binary alloys with no atomic size differences.
In general, we find that the minimum $R_c$ does not occur for
pure substances. For binary alloys containing atoms with different
crystalline symmetries, the minimum $R_c$ is
only a factor of $5$ lower than that for pure substances, which is
consistent with recent experimental studies of binary systems, such as
NiCu and ArKr, whose elements readily form FCC structures, as well as
high-entropy alloys. In contrast, $R_c$ for binary alloys
containing atoms with ICO and crystalline symmetries can be reduced by
three orders of magnitude relative to that for pure substances by
increasing $f_{\rm ICO}$.  These results emphasize that GFA of
binary alloys can be greatly
increased by mixing elements that enhance local icosahedral order
$f_i$.  However, $R_c(f_{i})$ is not monotonic; we show that $R_c$
possesses a minimum at a characteristic $f_{i} \gtrsim 0.075$, where
quasicrystals form. This result may explain why it is
difficult to obtain binary BMGs with large amounts of Al (since
it can lead to the formation of quasicrystals), whereas minor alloying
with Al can dramatically increase the GFA.

Although our results were obtained by studying binary alloys with
elements of the same size, they provide insights into the GFA of
alloys with elements of different sizes. For example, for CuZr, the
cohesive energies satisfy $\epsilon_{\rm Zr} > \epsilon_{\rm Cu}$, and
thus pure Cu (with FCC symmetry) is expected to have better GFA than
pure Zr (with HCP symmetry). (This result is confirmed by EAM
simulations in SM.) Further, Zr is larger than Cu with diameter ratio,
$\sigma_{\rm Cu}/\sigma_{\rm Zr} = 0.8$, and based on our prior
studies of binary Lennard-Jones systems~\cite{zhang_connection_2014},
Cu-rich alloys (with a majority of smaller atoms) have better
GFA. Thus, based on the cohesive energies and atomic sizes of Cu and
Zr, the composition with the best GFA should be Cu-rich. EAM
simulations for CuZr have shown that Cu$_{64}$Zr$_{36}$ is the
composition with the best GFA, and at this composition the local
icosohedral order is maximized~\cite{PRSI,ding_full_2014}. In future
studies, we will perform MD simulations of models of CuZr (and other
binary alloys) with effective pairwise interactions that include
cohesive energy and atomic size differences to identify the most promising 
BMG-forming binary alloys.
 
\section*{Acknowledgements}
The authors acknowledge support from NSF Grant Nos. DMR-1119826 (Y.-C.H.), CMMI-1901959 (C.O.), and CMMI-1463455 (M.S.). This work was supported by the High Performance Computing facilities operated by, and the staff of, the Yale Center for Research Computing. The authors thank P. Banner (Yale University), as well as S. Sarker and A. Mehta (SLAC National Accelerator Laboratory) for their contribution to the experimental studies.

\balance

%

\end{document}